\documentclass[aps,prb,10pt,twocolumn]{revtex4-2}
\usepackage{bm}
\usepackage{amsmath}
\usepackage{amsfonts}
\usepackage{amssymb}
\usepackage[colorlinks=true,linktocpage=true,breaklinks=true]{hyperref}

\usepackage{setspace}
\usepackage{extsizes}
\usepackage[version=3]{mhchem}
\usepackage[left=1.5cm, right=1.5cm, top=1.785cm, bottom=2.0cm]{geometry}
\usepackage{balance}
\usepackage{mathptmx}
\usepackage{graphicx} 
\usepackage{lastpage}
\usepackage[format=plain,justification=justified,singlelinecheck=false,font={stretch=1.125,small,sf},labelfont=bf,labelsep=space]{caption}
\usepackage{float}
\usepackage{fancyhdr}
\usepackage{fnpos}
\usepackage[english]{babel}
\addto{\captionsenglish}{%
  
}
\usepackage{array}
\usepackage{droidsans}
\usepackage{charter}
\usepackage[T1]{fontenc}
\usepackage[usenames,dvipsnames]{xcolor}
\usepackage{setspace}

\usepackage{tikz}
\usetikzlibrary{positioning}
\usepackage{subfigure}
\usepackage{amssymb}
\usepackage{flushend}

\usepackage{color,soul}

\usepackage{epstopdf}

\definecolor{cream}{RGB}{222,217,201}

\usepackage{listings}
\usepackage[toc]{appendix}

\usepackage{xcolor}

\definecolor{codegreen}{rgb}{0,0.6,0}
\definecolor{codegray}{rgb}{0.5,0.5,0.5}
\definecolor{codepurple}{rgb}{0.58,0,0.82}
\definecolor{backcolour}{rgb}{0.95,0.95,0.92}

\lstdefinestyle{mystyle}{
    backgroundcolor=\color{backcolour},   
    commentstyle=\color{codegreen},
    keywordstyle=\color{magenta},
    numberstyle=\tiny\color{codegray},
    stringstyle=\color{codepurple},
    basicstyle=\ttfamily\footnotesize,
    breakatwhitespace=false,         
    breaklines=true,                 
    captionpos=b,                    
    keepspaces=true,                 
    numbers=left,                    
    numbersep=10pt,                  
    showspaces=false,                
    showstringspaces=false,
    showtabs=false,                  
    tabsize=4
}

\begin{document}

\title{Classical and quantum machine learning applications in  spintronics}

\author{Kumar J. B. Ghosh}
\email{jb.ghosh@outlook.com}
\affiliation{E.ON Digital Technology GmbH, 
45131, Essen, Germany.}

\author{Sumit Ghosh}
\email{s.ghosh@fz-juelich.de}
\affiliation{Institute of Advance Simulations, Forschungszentrum J{\"u}lich GmbH, 52428 J{\"u}lich, Germany}
\affiliation{Institute of Physics, Johannes Gutenberg-University Mainz, 55128 Mainz, Germany}

\begin{abstract}
In this article we demonstrate the applications of classical and quantum machine learning in quantum transport and spintronics. With the help of a two-terminal device with magnetic impurity we show how machine learning algorithms can predict the highly non-linear nature of conductance as well as the non-equilibrium spin response function for any random magnetic configuration.  
By mapping this quantum mechanical problem onto a classification problem, we are able to obtain much higher accuracy beyond the linear response regime  compared to the prediction obtained with conventional regression methods.  
We finally describe the applicability of quantum machine learning which has the capability to handle a significantly large configuration space. Our approach is  applicable for solid state devices as well as for molecular systems. These outcomes are crucial in predicting the behavior of large-scale systems where a quantum mechanical calculation is computationally challenging and therefore would play a crucial role in designing nano devices.
\end{abstract}

\maketitle

\section{Introduction} \label{sec: introduction}

In recent years machine learning techniques \cite{russell2003artificial} have become powerful tools in various research fields, for e.g., material science and chemistry \cite{artrith2021best, schutt2019unifying, janet2020machine}, power and energy sector \cite{b23, ghoddusi2019machine}, cyber security and anomaly detection\cite{8359287, Omar2013MachineLT}, drug discovery \cite{vamathevan2019applications}, etc.  
These techniques can be implemented on classical as well as quantum computers \cite{nielsen2002quantum} which makes them even more powerful specially for problems which are unsolvable by any conventional means. There are extensive ongoing efforts on the application of quantum computing in the areas of machine learning \cite{schuld2015introduction, biamonte2017quantum, sakhnenko2021hybrid}, finance \cite{woerner2019quantum},  quantum chemistry \cite{lanyon2010towards,cao2019quantum}, drug design and molecular modeling \cite{kandala2017hardware}, power systems \cite{ghoshIEEE, eskandarpour2021experimental}, metrology \cite{giovannetti2004quantum}, to name a few applications. Quantum-enabled methods are the next natural step of the AI studies to support faster computation and more accurate decision making, creating the interdisciplinary field of quantum artificial intelligence \cite{wichert2020principles}.

Recently machine learning (ML) and quantum computing (QC) applications are gaining attention in the field of condensed matter physics \cite{bedolla2020machine,  xia2018quantum, weber2010quantum, smith2019simulating}. Most of the studies so far are focused on the electronic properties \cite{chandrasekaran2019solving, Westermayr2021, Fiedler2022} or transport properties \cite{Lopez2014, Wu2020}. The application of ML has significantly reduced the computational requirement as well as time consumption for computationally demanding problems. In this paper we address another very active and promising brunch of condensed matter physics - namely $spintronics$ which is focused on manipulating spin degree of freedom and has been in the heart of modern computational device technology. Here we employ classical and quantum machine learning algorithm to predict two main observables in spintronics, namely non-equilibrium spin density generated by an applied electric field and the transmission coefficient  in a two terminal device configuration in presence of magnetic impurity. This configuration is the basis of any magnetic memory device where the non-equilibrium spin density provides the torque necessary for manipulating the magnetization \cite{Manchon2008, Manchon2009}. The theoretical evaluation of non-equilibrium spin density is done via non-equilibrium Green's function technique \cite{Ghosh2017, Ghosh2018, Ghosh2019} which is computationally quite demanding. Compared to that, prediction with trained learning algorithm is quite efficient \cite{Lopez2014, Wu2020}  and allows to study a large number of configurations. For a given system, the spintronic properties are usually dominated by a subset of parameters necessary to define the whole system. In this machine learning approach only a limited number of parameters are used to construct the feature space; therefore, the dimensionality of the problem is significantly reduced. In our case we chose the magnetization configuration and the transport energy as the governing parameters. For a given arbitrary distribution of magnetization, the spin response functions as well as the transmission coefficient can be a highly non-linear function of the transport energy. For such high level of non-linearity, conventional regression methods fail to provide reliable outcome over a broad energy range. In this paper we present a new approach to handle this problem. By discretizing the continuous outcome, we convert the nonlinear regression into a classification problem and obtained a high level of accuracy with a classical machine learning algorithm.  We systematically analyzed the transmission and the spin response function over a large range of transport energies and internal parameters. Finally, we also demonstrate the applicability of quantum machine learning algorithm which can be useful for exponentially large configuration that is beyond the scope of any classical algorithm.

The organization of this article is as follows. After a brief introduction in Sec.\,\ref{sec: introduction}, we define our model and methods in Sec.\,\ref{sec: Model and method}. It contains the non-equilibrium Green's function method used to generate the training data as well as the classical and quantum ML approach along with our discretization scheme used to analyze the data. The results and discussions are given in Sec.\,\ref{sec: Results and Discussions}, which contains the outcomes of both classical and quantum ML. Finally, in Sec.\,\ref{sec: Conclusions}, we offer some concluding remarks.

\section{Model and method} \label{sec: Model and method}

\subsection{Tight binding model and non-equilibrium Green's function approach}

\begin{figure}[ht!]
\centering
\includegraphics[width=0.48\textwidth]{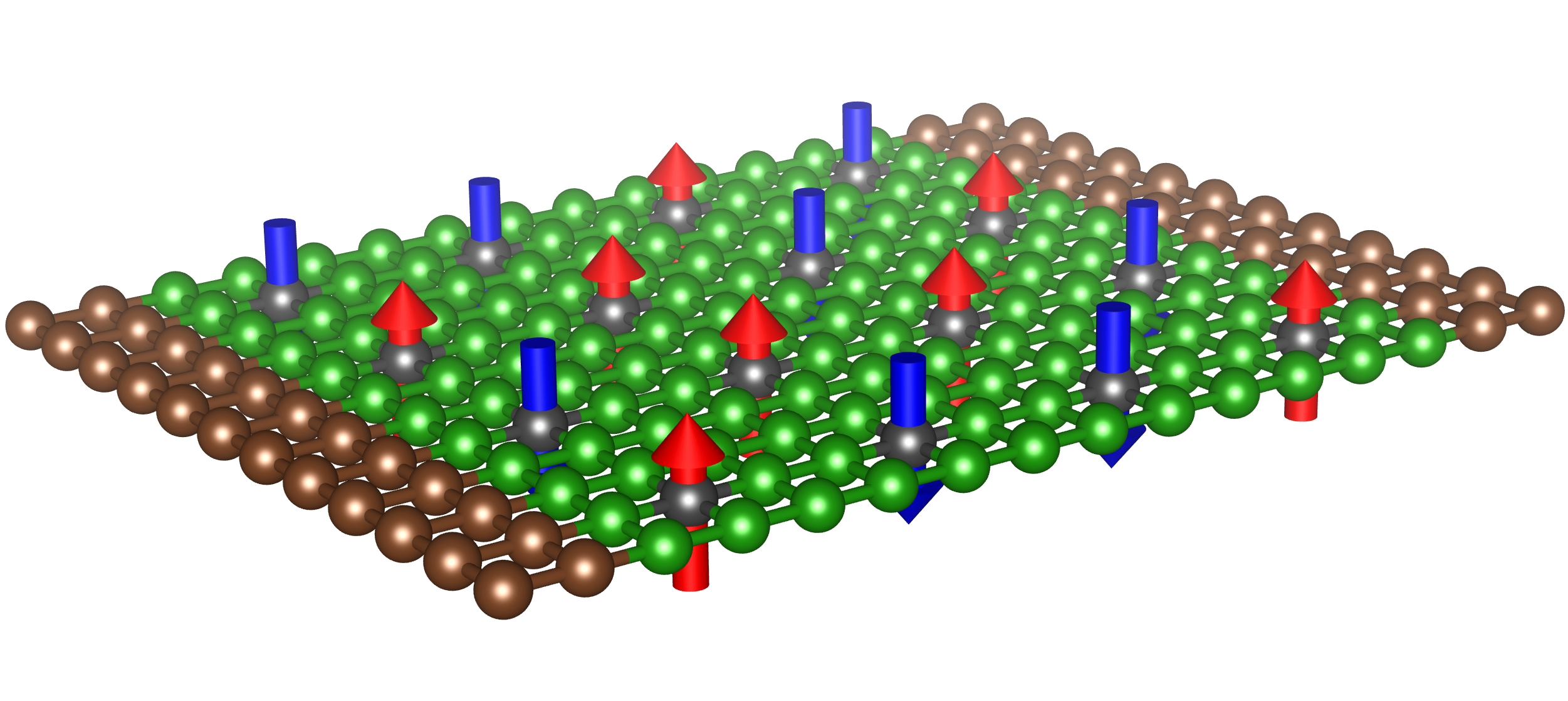}
\caption{Schematic of a two-terminal device. Green region shows the scattering region. The green sites show the non-magnetic sites and gray sites show magnetic sites with up (red) and down (blue) magnetization. }
\label{fig1}
\end{figure}

In this study, we use a two terminal device configuration where a scattering region with magnetic impurity is attached to two semi-infinite non-magnetic electrodes. Here we use only out of plane magnetization, however this formalism is also applicable for the non-collinear magnetization as well. The system is defined with a tight binding Hamiltonian 

\begin{eqnarray}
H &=& \sum_{i,\mu \nu} c^\dagger_{i,\mu} \epsilon_{i}^{\mu\nu} c_{i,\nu} + \sum_{\langle i j \rangle, \mu \nu} c^\dagger_{i,\mu} t_{ij}^{\mu\nu} c_{j,\nu}
\end{eqnarray}

where $\epsilon_i^{\mu \nu}$ is the onsite potential and $t_{ij}^{\mu\nu}$ is the nearest neighbor hopping term. Here we consider Rashba-Bychkov type hopping for the spin dependent part which can be realized on the surface of a heavy metal such as Pt or W and can be induced to other material with proximity effect. The full hopping term along $\hat{x}$ and $\hat{y}$ direction is given by 
\begin{equation}
t_{\bm{r}+\hat{x}}= t_0 \mathbb{I}_2-i t_R \sigma_y, ~~ t_{\bm{r}+\hat{y}} = t_0 \mathbb{I}_2+i t_R \sigma_x ,
\end{equation}
where $\mathbb{I}_2$ is the identity matrix of rank 2 and $\sigma_{x,y,z}$ are the Pauli matrices. $t_0$ is the spin independent hopping amplitude and $t_R$ is the Rashba coefficient. The onsite energies also consist of both magnetic and nonmagentic parts and is given by 
\begin{equation}
\epsilon_{i}= -4t_0 \mathbb{I}_2 + m_i \Delta \sigma_z ,
\end{equation}
where $m_i$=$0,\pm 1$ corresponding to non-magnetic sites, sites with positive and negative magnetization respectively,  and $\Delta$ is the exchange energy. We choose the exchange energy $\Delta$ as the unit of our energy and choose $t_0$=-0.5$\Delta$. Unless otherwise mentioned $t_R$ is kept at 0.1$\Delta$. We consider a $12 \times 12$ scattering region with uniformly spaced 16 magnetic centers (Fig.\ref{fig1}) where the magnetization directions are being chosen randomly. The electrodes are chosen to be non-magnetic with same hopping parameters

The conductance of the system is calculated using Green's function. For simplicity we adopt natural unit here ($c$=$e$=$\hbar$=1). The transmission probability and therefore the conductance from left to right electrode is given by 
\begin{equation}
T = Tr\left[\Gamma_1 G^R \Gamma_2 G^A\right] ,
\end{equation}
where 
\begin{equation}
G^{R,A}=\left[E-H_S-\Sigma^{R,A}_1-\Sigma^{R,A}_2\right]^{-1}
\end{equation}
is the retarded/advanced Green's function of the scattering region, and 
\begin{equation}
\Gamma_{1,2}=i\left[\Sigma_{1,2}^R - \Sigma_{1,2}^A\right],
\end{equation}
with $\Sigma_{1,2}^{R,A}$ being the retarded/advanced self energy of the left/right electrode. To calculate the non-equilibrium spin densities one can utilize the lesser Green's function \cite{Nikolic2005, Nikolic2018} defined as 
\begin{equation}
G^<(E)= G^R(E) \Sigma^<(E) G^A(E) ,
\end{equation}
where
\begin{equation}
\Sigma^<(E)=i\left[f_1(E)\Gamma_1(E)+f_2(E)\Gamma_2(E)\right],
\end{equation}
with $f(E)$ being the Fermi-Dirac distribution of the corresponding electrode. The non-equilibrium expectation value of an observable $\mathcal{O}$ at energy $E$ subjected to a bias voltage $V$ is given by
\begin{eqnarray}
\langle \hat{\mathcal{O}} \rangle_E = \int_{E-V/2}^{E+V/2} d\epsilon ~ Tr \left[ \mathcal{\hat{O}} \cdot \rho(\epsilon) \right] ,
\end{eqnarray}
where $\rho(E)$=$\frac{1}{2\pi i}G^<(E)$ is the non-equilibrium density matrix. For an infinitesimal bias voltage ($V \to 0$) it is convenient to calculate the response function. Here we are interested in the response function for the in-plane spin component given by 
\begin{equation}
S_i^{x,y} = Tr \left[\sigma_{x,y} \cdot \rho_i \right] ,
\end{equation}
where $\rho_i$ being the projection of the non-equilibrium density matrix on $i$th site. For our calculation we use the tight-binding software KWANT \cite{Groth2014} where the non-equilibrium density matrix can be obtained via the scattering wave-function. We generate the conductance and in-plane spin response for randomly chosen spin configurations and energies and use them to train our algorithm. 


\subsection{Non-linearity of the response} 

\begin{figure}[ht!]
\centering
\includegraphics[width=0.48\textwidth]{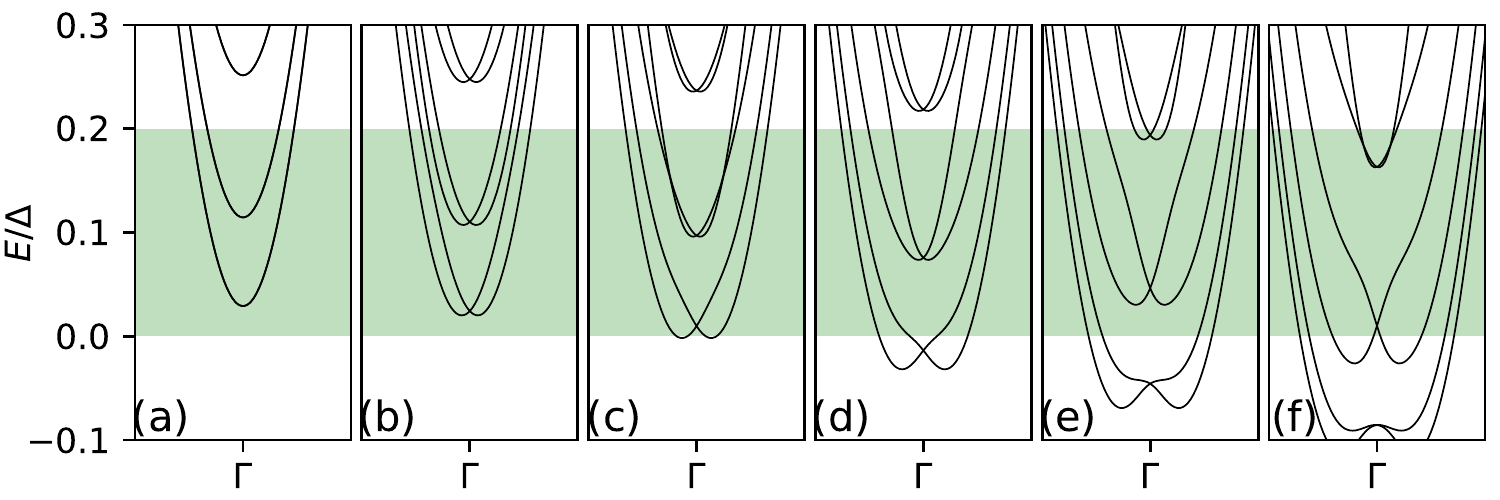}
\caption{Variation of lead band structure with $t_R$. (a), (b), (c), (d), (e), and (f) show the band structures for $t_R$=0.00$\Delta$, 0.05$\Delta$, 0.10$\Delta$, 0.15$\Delta$, 0.20$\Delta$, and 0.25$\Delta$ respectively. The green region denotes the energy window where the analysis has been done.}
\label{bands}
\end{figure}

Let us first consider the intrinsic nature of the system under consideration and the inherent non-linearity of its conductance and spin response function. We start by looking at the band structures of the non-magnetic electrodes for different values of $t_R$ (Fig.\,\ref{bands}).

\begin{figure}[ht!]
\centering
\includegraphics[width=0.48\textwidth]{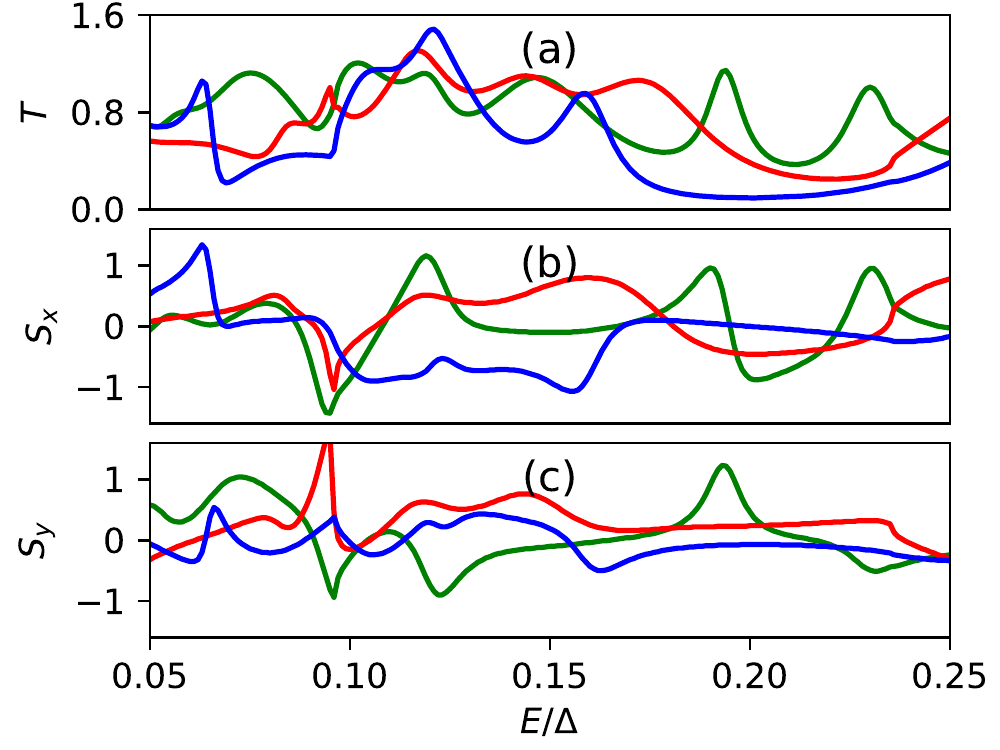}
\caption{Variation of (a) conductance ($T$) and spin response, (b) $S_x$, and (c) $S_y$ function on 6th magnetic site.}
\label{txy}
\end{figure}

For a clean and homogeneous system, the transmission probability and therefore the conductance shows a step like behaviour. In presence of the magnetic sites in the scattering region, this behaviour becomes highly non linear. For this study we focus on three different entities, namely the conductance($T$) and the $x$ and $y$ component of the spin response on the magnetic sites (Fig.\,\ref{txy}).

One can readily see from Fig.\ref{txy} that the responses are highly nonlinear in nature within our chosen energy window and completely uncorrelated for different magnetic configuration. 
For simplicity we consider collinear magnetism ($m_i=\pm 1$) while the energy is kept as a continuous variable. The formalism is also applicable for non-collinear magnetism, however it would expand the input parameter space since each magnetic moment has to be described by three components.

\subsection{Classical and Quantum machine learning} \label{subsec: Classical and Quantum machine learning}
Any machine learning approach consists of two steps - training and testing. For training one has to consider a large number of data where both inputs and outputs are known. For testing we use new input values and predict the output. For our case, we consider 17 input parameters . First 16 are the magnetization direction of the 16 magnetic sites denoted by integers (1 for $\uparrow$ spin and -1 for $\downarrow$ spin) and the 17th input is the energy at which we calculate the desired output and is a floating number between 0.0 and 0.2. 
For output we consider conductance of the system and the $x$ and $y$ components of non-equilibrium spin density at each of the $16$ magnetic sites. The sample data is produced using non-equilibrium Green's function method which is computationally quite demanding since it requires quantum mechanical description of the complete system including the non-magnetic sites and the electrodes. Depending on the method and observable of interest these calculations can scale an $h^3$ or at best $h$ where $h$ is the dimension of the Hamiltonian matrix of the complete system. Here we choose our system large enough to demonstrate the significant non-linearity in the physical observables. The machine learning approach we present here, however, is not restricted by the dimension of the physical system.

First we compare the performance of different classification algorithms, for e.g., Logistic regression \cite{kleinbaum2002logistic}, k-nearest neighbors (KNN) \cite{Kramer2013}, Random Forest \cite{breiman2001random}, Support Vector Machine (SVM) \cite{b29}, etc. to train the models. Then, we use the trained models on the respective test samples and obtain the outputs. Among all the above classifiers the Random Forest performs the best and therefore we consider Random Forest throughout this paper. 
For comparison we also choose different regression models, for e.g., Theil-Sen regressor \cite{theil1950rank, Sen_1968}, RANSAC (Random Sample Consensus) regressor \cite{fischler1981random}, and SGD (Stochastic Gradient Descent) regressor \cite{Bottou2007} for the data analysis, but the regressors perform much worse than the classifiers.

For a complex inhomogeneous multilevel nano-device the number of governing parameters can be exponentially large which can be challenging for a classical computer. For such cases quantum machine learning algorithms can provide an efficient alternative. One of the most popular quantum classifiers is  Quantum Support Vector Machine (QSVM) \cite{b32,b33}, which is a quantized version classical SVM \cite{b29}.  It performs the SVM algorithm using quantum computers. It calculates the  kernel-matrix using the quantum algorithm for the inner product on quantum random access memory (QRAM) \cite{giovannetti2008quantum}, and performs the classification of query data using the trained qubits with a quantum algorithm.  The overall complexity of the quantum SVM is $\mathcal{O}\left(\log (NM)\right)$, whereas classical complexity of the SVM is $\mathcal{O}\left({M}^{2} ~(M+N)\right)$, where $N$ is the dimension of the feature space and $M$ is the number of training vectors. The complexity of the Random forest (the best performing algorithm for our dataset) is $\mathcal{O}\left(T N M log M \right)$, where $M$, $N$, and $T$ are the number of instances in the training data, the number of attributes, and the number of trees respectively. Therefore, the QSVM model for the solution of classification and prediction offers upto exponential speed-up over its classical counterpart. Beside QSVM, an alternate class of quantum  classification algorithm is introduced \cite{b34, b37}, called Variational Quantum  Classifier (VQC). This NISQ-friendly algorithm operates through using a variational quantum circuit to classify a training set in direct analogy to the conventional SVMs. 


\subsection{Regression vs classification}

\begin{figure}[ht!]
\centering
\includegraphics[width=0.48\textwidth]{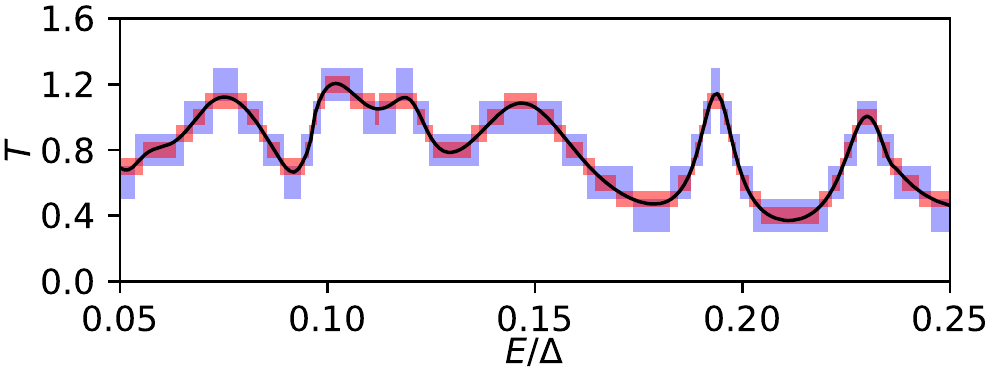}
\caption{Discretisation of the continuous output. Blue and red boxes correspond a block height of 0.2 and 0.1.}
\label{disct}
\end{figure}

Convetionally, the physical observables are calculated within linear response regime where linear regression can provide reasonable accuracy \cite{Wu2020}. However, for a highly non-linear response, such as Fig.\,\ref{txy}, applicability of regression becomes quite non-trivial. To increase the accuracy and efficiency of the learning process, here we adopt an alternative approach. First we discretise the output within small blocks and assign a class to each block (Fig.\,\ref{disct}). To demonstrate that we consider the transmission spectrum corresponding to the green line in Fig.\,\ref{txy}.

For a block height $\delta$, the class of an output $y$ is defined as $\mathcal{C}$=Round$[y/\delta]$, where Round[] represents rounding off to the nearest integer. In this way a trained network can predict a class $\mathcal{C}$ for a unknown set of input parameters, from which one can retrieve the actual value $y$ as $y$=$\mathcal{C}\delta$. Therefore $\delta$ correspond the intrinsic uncertainty of the discretisation.  A larger value of $\delta$ would reduce the number of classes and therefore increase the accuracy of the prediction, however the predicted value can significantly differ from the actual value due to the uncertainty posed by $\delta$ and therefore increase the overall error. A small value of $\delta$ on the other hand can reduce the uncertainty, however it would increase the number of classes significantly and therefore may pose a computational challenge for the learning algorithm.

\section{Results and Discussions} \label{sec: Results and Discussions}

\begin{figure*}[htb]
\centering
\captionsetup{justification=centering,margin=2cm}
\includegraphics[width=0.9\textwidth]{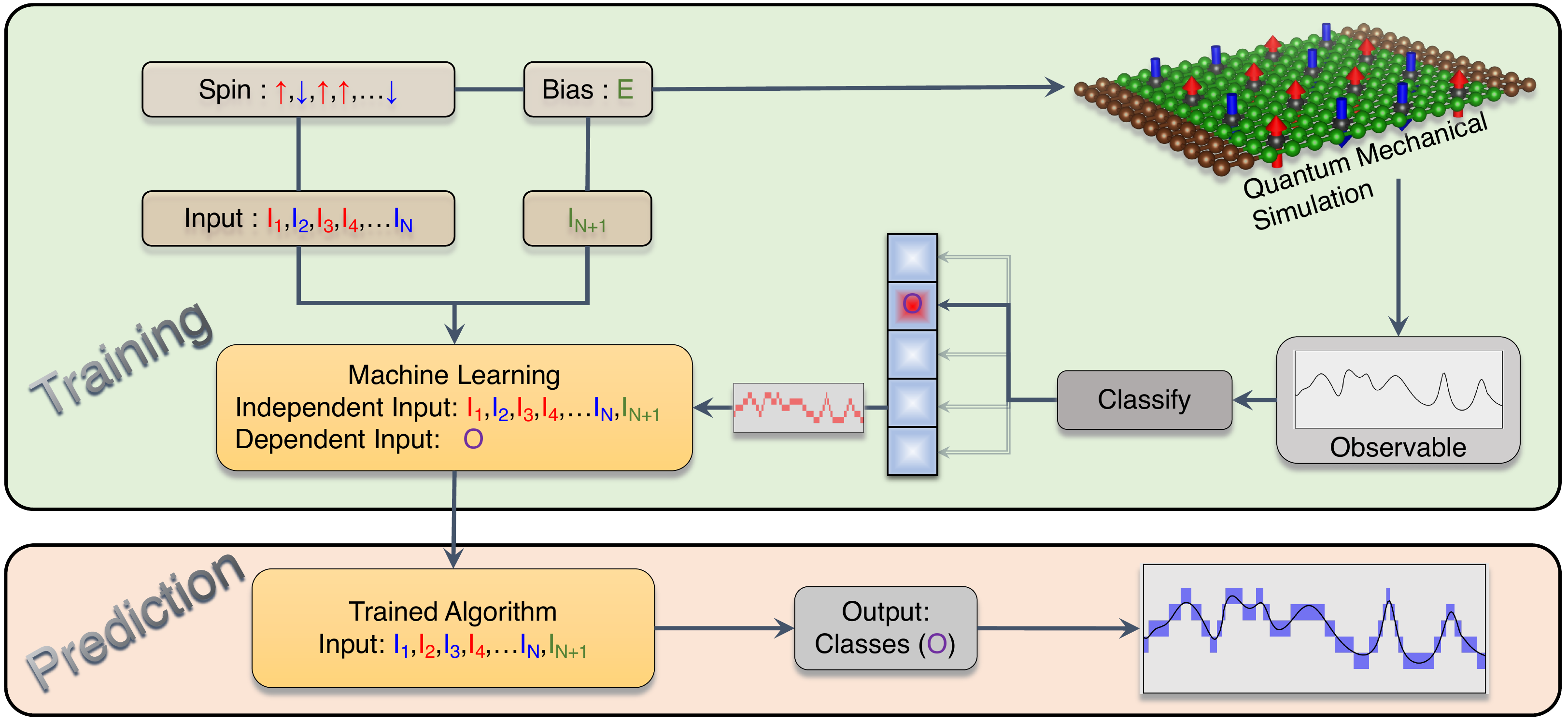}
\caption{Schematic representation of the data analysis.}
\label{Schematic_diagram}
\end{figure*}

As mentioned in  Sec.\,\ref{sec: Model and method}, we consider a scattering region with 16 magnetic sites where the magnetizations can either point up or down. This gives a total of $2^{16}$ different configurations. For each of this configurations, one can calculate the transmission at any arbitrary energy which we choose between 0-0.2$\Delta$. We are therefore dealing with a 17 dimensional feature space with mixed input variables where the first 16 inputs are either -1 (for spin $\downarrow$) or 1 (for spin $\uparrow$) and the 17th input is a floating number between 0 and 200 denoting the energy. For our study, we consider a set of $10^5$ random input configurations and calculate corresponding transmission values and both the $x$ and $y$ component of spin response functions on all 16 magnetic sites. The theoretical workflow has been outlined in Fig.\ref{Schematic_diagram}.

It is worth mentioning that the state-of-the art AI models can handle billions of parameters which requires months of training. However, for most physical problem the challenge is to express the physical observable as a function of minimum number of parameters. Besides, experimentally one can obtain only few features of a system and therefore for practical use one requires a method which can predict a highly nonlinear outcome from fewer input parameter which is the main objective of this work.

\subsection{Success rate vs accuracy with number of classes}

\begin{figure}[t!]
\centering
\includegraphics[width=0.48\textwidth]{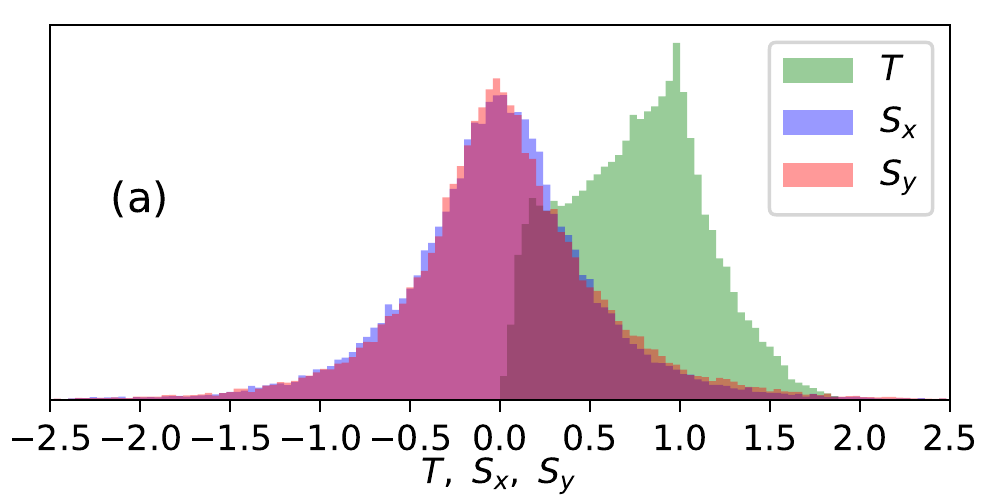}
\includegraphics[width=0.48\textwidth]{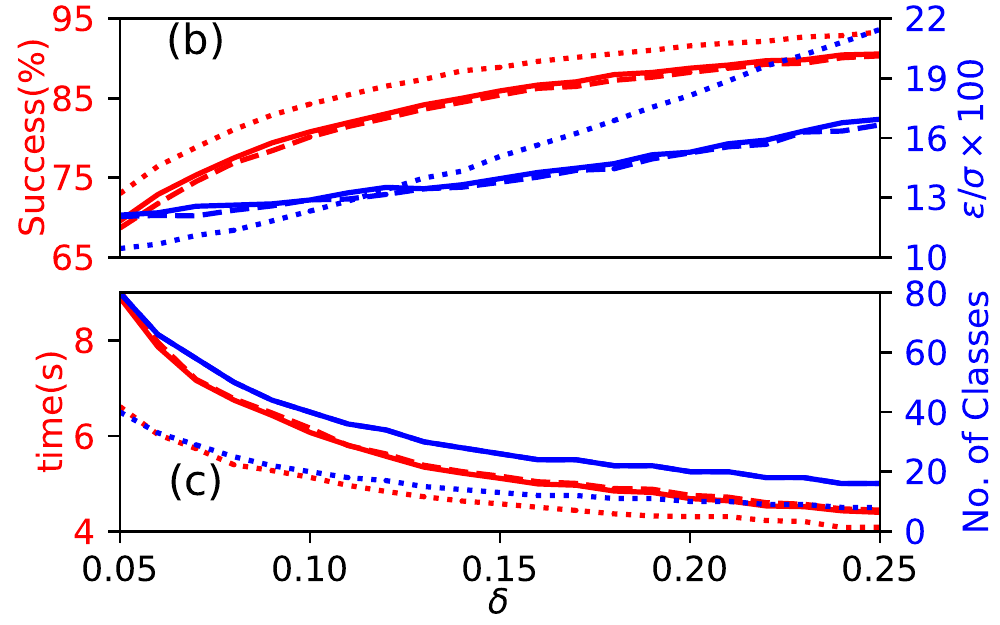}
\caption{Comparison of predictions for $T$, $S_x$, and $S_y$ with respect to the discretization parameter $\delta$. (a) Distribution of the values of $T$, $S_x$, and $S_y$ for $\delta=0.1$. (b) Success rate of the prediction (red) and accuracy ($\epsilon/\sigma$) (blue), where the solid, dashed, and dotted lines correspond to $S_x$, $S_y$, and $T$ respectively. (c) Time consumption (red) and number of classes (blue) for $S_x$ (solid) and $S_y$ (dashed) and $T$ (dotted).}
\label{hist}
\end{figure}

The samples are randomly split into $9\times10^4$ training data and $10^4$ testing data and then we conduct 50 different  train-test cycles. 
The number of classes depends on the choice of the parameter $\delta$. As discussed earlier, reducing $\delta$ can decrease the error, however it also increases the number of classes and therefore reduce the accuracy. Unless otherwise mentioned, we keep $\delta$=0.1 which provides good balance between accuracy and error. Due to the highly non-linear nature of the system, there are few high values of the physical observable (Fig.\,\ref{hist}a) which can significantly increase the total number of classes where the higher classes would have insignificant population. This in turns can reduce the performance of the learning algorithm. To avoid this scenario we put an upper cutoff of 2 for $T$ and $S_{x,y}$, which means any value greater/less than $\pm$2 is considered as $\pm$2.
The performance of prediction is characterized in terms of the success rate and accuracy, where the accuracy is defined as the ratio of the root mean square error (RMSE) $\epsilon$  of the prediction to the standard deviation $\sigma$ of the training data (Fig.\ref{hist}a).  This scales down the change of accuracy due to the variation of distribution of output classes. We try several training algorithms  such as KNeighbors, DecisionTree and RandomForest. 
Among these methods Random forest shows better performance within reasonable execution time, and therefore we  use Random forest throughout the rest of the study.

Note that unlike $T$, $S_{x,y}$ can have both positive and negative values and therefore for the same value of $\delta$ results in twice the number of classes for $S_{x,y}$ compared to $T$ (Fig.\ref{hist}c). This enhancement of classes along with the localization of spin density, as shown by  the peaks in  cause a slight reduction of success rate and detection efficiency compared to that of $T$ (Fig.\,\ref{hist}b).

\subsection{Prediction of transmission and spin response functions}

As one can see from Fig.\,\ref{bands}, the band structure and therefore the physical properties depend crucially on the choice of parameter. This in turns affects the distribution of the outputs and therefore the prediction itself. To demonstrate that we consider six different values of the parameter $t_R$, as showed in Fig.\,\ref{bands} and calculate $10^5$ sample points by randomly varying the onsite magnetization $m_i$ and energy where the energy values are kept within $[0,0.2\Delta]$. Training is done with randomly chose $9\times10^4$ data and the testing is done on rest of the $10^4$ data points using Random forest algorithm. The success and accuracy are calculated by averaging over $50$ different train-test cycles. The $50$-fold cross validation ensures the model to be free from overfitting.

\begin{table}[ht!]
\centering
\begin{tabular}{|c|c|c|c|c|c|}
\hline
$t_R/\Delta$ & Success($\%$) & $\epsilon/\sigma$  & $N_{class}$ & $t_{Train}$(s) & $t_{Test}$(s) \\ \hline
0.00 & 85.90 & 13.94 & 25 & 5.06 & 0.20 \\ \hline
0.05 & 84.46 & 12.41 & 22 & 5.15 & 0.20 \\ \hline
0.10 & 84.33 & 12.28 & 21 & 5.26 & 0.21 \\ \hline
0.15 & 87.50 & 10.63 & 22 & 4.97 & 0.20 \\ \hline
0.20 & 89.20 & 11.80 & 19 & 4.80 & 0.19 \\ \hline
0.25 & 90.46 &  9.82 & 20 & 4.78 & 0.19 \\ \hline
\end{tabular}
\caption{Qualitative variation of the prediction with respect to the Rashba parameter $t_R$.}
\label{tab:tr}
\end{table}

\begin{figure}[ht!]
\centering
\includegraphics[width=0.48\textwidth]{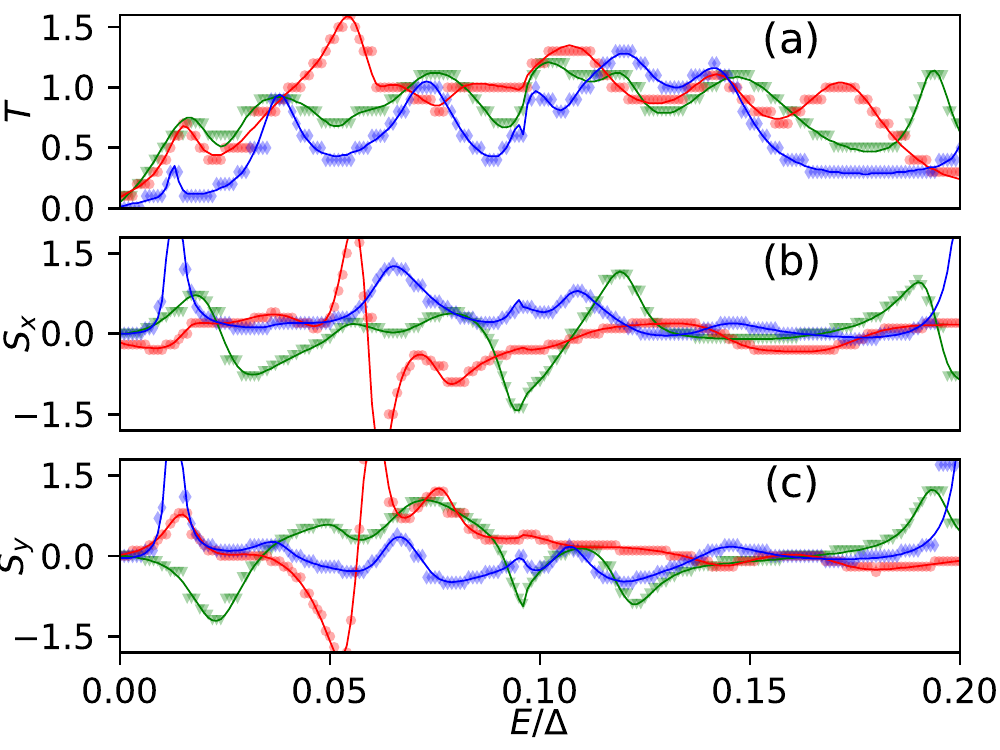}
\caption{Comparison of the predicted values against the actual values of (a) $T$,  (b) $S_x$, and (c) $S_y$ for three different configurations. The symbols  show the predicted values and the lines show the numerically calculated  values (Fig.\,\ref{txy}).}
\label{txyfit}
\end{figure}

From Table\,\ref{tab:tr}, one can see that the quality of prediction gets better for higher value of $t_R$. This is because for smaller values of $t_R$, the entire energy range (green region in Fig.\,\ref{bands}) is not spanned by bands and therefore for a large number of input data the output remains 0. As we increase the value of $t_R$ the selected energy range is covered with bands resulting more ordered finite output. 
Physically, an increased Rashba parameter can suppress scattering and therefore reducing the fluctuation of the transmission which result in a better prediction.
For rest of the paper we consider $t_R$=0.1$\Delta$. To demonstrate the quality of the prediction we consider three configurations showed in Fig.\,\ref{txy}a and evaluate the transmission coefficient on uniformly spaced energy values (Fig.\ref{txyfit}a).

In our test system we have 16 magnetic centers where we calculate the spin response function. For this study we keep $t_R$=0.1$\Delta$ and train with Random forest algorithm. For brevity, we show $S_x$ and $S_y$ only at 6th magnetic site which has been showed for three specific configurations in Fig.\,\ref{txy}.  To demonstrate the quality of our prediction we also consider three particular configurations (Figs.\,\ref{txy}b, \ref{txy}c) and showed the predicted values against calculated values (Figs.\,\ref{txyfit}b, \ref{txyfit}c).  


\subsection{Application of quantum classifier}\label{subsec: Application of quantum classifier}

\begin{figure}[ht!]
\centering
\includegraphics[width=0.48\textwidth]{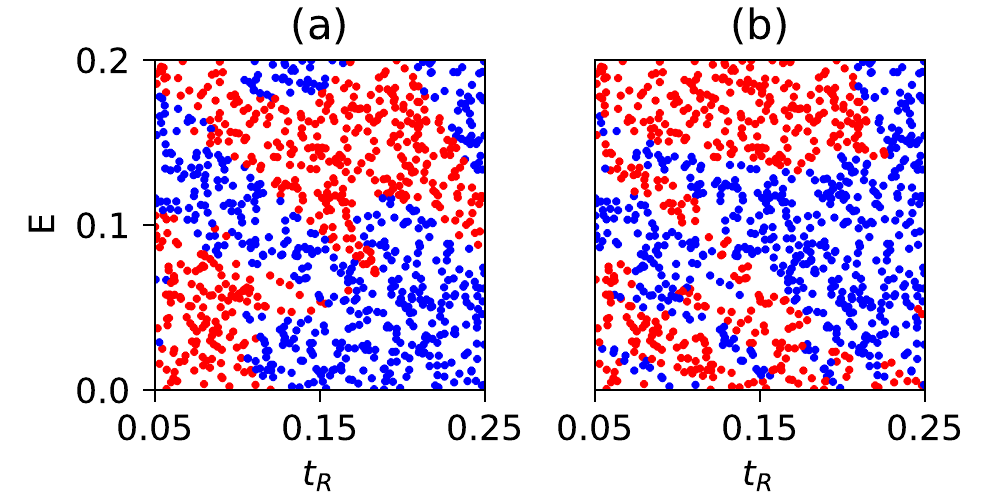}
\caption{A sample dataset with two features and two classes. Blue and red dots show the $0$ and $1$ classes for (a) $S_x^6$ and (b) $S_y^6$}
\label{fig6}
\end{figure}

Finally we demonstrate the feasibility of quantum machine learning (QML) for our problem. Due to limitation of resources it is not possible to handle large number of input parameter or classes in this case. Therefore, we consider a particular magnetic configuration and choose the Rashba parameter ($t_R$) and the transmission energy ($E$) for the two components of the input variable and the sign of non-equilibrium $S_{x,y}$ on each site as the two output classes. Physically speaking the sign of $S_x$ and $S_y$ determines the switching direction and direction of precession of the magnetic moments. We generate 1000 random input point in this two dimensional $t_R$-$E$ space and evaluate the sign of $S_{x,y}$ for each of the 16 magnetic sites. A sample dataset is presented in Fig. \ref{fig6}. 

We divide each dataset into two parts, namely, training data ($900$ data points) and testing data ($100$ data points). We implement  classical SVM using Scikit-learn \cite{pedregosa2011scikit}, and  QSVM with Qiskit \cite{Qiskit} from IBMQ, using different feature maps (for e.g., ZFeatureMap, ZZFeatureMap, etc.), to classify the data.
We repeat the above procedure with all the $16$ datasets and summarize the result in table \ref{table_qsvm}. For brevity we show $S_{x,y}$ for only first 8 sites.
\begin{table}[ht]
    \centering
    \begin{tabular}{|c|c|c|c|c|}
    \hline
        Quantity & QSVM & SVM(RBF) & SVM(Lin) & SVM(Poly) \\ \hline
        $S_x^1$ &  83\% & 81\% & 58\% & 58\% \\ \hline
        $S_y^1$ &  78\% & 77\% & 71\% & 71\% \\ \hline
        $S_x^2$ &  83\% & 80\% & 79\% & 79\% \\ \hline
        $S_y^2$ &  90\% & 92\% & 93\% & 93\% \\ \hline
        $S_x^3$ &  77\% & 69\% & 54\% & 64\% \\ \hline
        $S_y^3$ &  79\% & 75\% & 62\% & 71\% \\ \hline
        $S_x^4$ &  85\% & 76\% & 71\% & 68\% \\ \hline
        $S_y^5$ &  82\% & 79\% & 82\% & 78\% \\ \hline
        $S_x^5$ &  82\% & 78\% & 73\% & 73\% \\ \hline
        $S_y^5$ &  84\% & 84\% & 51\% & 64\% \\ \hline
        $S_x^6$ &  83\% & 89\% & 64\% & 67\% \\ \hline
        $S_y^6$ &  76\% & 75\% & 63\% & 69\% \\ \hline
        $S_x^7$ &  75\% & 75\% & 58\% & 70\% \\ \hline
        $S_y^7$ &  80\% & 73\% & 63\% & 70\% \\ \hline
        $S_x^8$ &  78\% & 81\% & 66\% & 68\% \\ \hline
        $S_y^8$ &  74\% & 75\% & 60\% & 64\% \\ \hline
    \end{tabular}
    \caption{Comparing the testing accuracies between different classical and quantum classifiers for the $S_x$ and $S_y$ for first 8 magnetic sites. In the above table RBF, Lin, and Poly represents the RBF, Linear, and Polynomial Kernels used in SVM algorithm.}
    \label{table_qsvm}
\end{table}

From table \ref{table_qsvm}, we see that the quantum classifier is performing better than its classical counterparts in many cases.  Although, the main advantage of QML over classical ML is in the runtime  (see Sec.\,\ref{subsec: Classical and Quantum machine learning}), so that, for a significantly larger data size and configuration space QML will be the only feasible option. Therefore, with the availability of sufficient quantum computing resources this approach will be very useful to analyse the large solid state and molecular devices as well.

\section{Conclusion} \label{sec: Conclusions}

In this article, we demonstrate the applicability of different classical and quantum machine learning approachs for spintronics. We show that how one can achieve a significantly improved performance by converting the conventional regression problem into a discretised classification problem. Our approach allows us to obtain a high level of accuracy even for a strongly nonlinear regime. We further demonstrate the applicability of quantum machine learning which performs quite well for our small feature space. Considering the scalability of quantum machine learning algorithms over their classical counter parts (see Sec.\,\ref{subsec: Classical and Quantum machine learning}) this will significantly enhance the performance for a larger configuration space and data size; in fact QML will be the only  viable option in that regime. Our method is quite generic and therefore is equally applicable for a large class of systems, especially, for the molecular devices where one can easily use charge or orbital degrees of freedom along with the spin to control different physical observables. Our work thus opens new possibilities to study a large variety of physical system and their physical properties with machine learning.

\bibliographystyle{apsrev4-2}
\bibliography{ref}

\section{Relevant codes for data analysis and machine learning}\label{appendix: A}

The supporting data and codes for this study are available in the following \href{https://github.com/jbghosh/ML_QML_Spintronics}{GitHub repository}. Classical ML is implemented in $RF\_fit.ipynb$. $Train.npy$ contains $10^5$ training data and $Test.npy$ contains testing data for three specific configurations used in Fig.\ref{txy}, where each configuration has 201 uniformly spaced energy values. 
The data structure is as follows: first 16 columns describe the magnetic configurations of the 16 magnetic sites. 17th column represents the energy at which the desired output is computed. 18th column denotes the transmission. 19th and 20th column are the spin-components  for the 6th magnetic site respectively. A sample code for classical data analysis is described below.

\begin{lstlisting}[language=Python]
import numpy as np
from sklearn.model_selection import train_test_split
import matplotlib.pyplot as plt
from sklearn.metrics import accuracy_score
from sklearn.ensemble import RandomForestClassifier

train_data=np.load("Train.npy")
test_data=np.load("Test.npy")

def data_analysis(ncol,train,test,dy):
    x_train=train[:,0:17]; y_train=train[:,ncol]
    x_test = test[:,0:17]; y_test = test[:,ncol]
    y_train[y_train > 2.0]= 2.0
    y_train[y_train <-2.0]=-2.0
    y_test[y_test > 2.0]= 2.0
    y_test[y_test <-2.0]=-2.0

    #Convert values into class
    y_train_cl=np.rint(y_train/dy)
    y_test_cl =np.rint(y_test/dy)

    #classification with random forest
    clf = RandomForestClassifier(n_jobs=None)
    clf.fit(x_train,y_train_cl)
    y_pred_rf = clf.predict(x_test)
    acq = clf.score(x_test,y_test_cl)
    ydat=np.stack((x_test[:,-1], y_test, y_pred_rf*dy),axis=-1)
    np.save("col%s.npy"%ncol,ydat)

    #Testing with 10% of training data
    x_data=x_train
    y_data=y_train_cl
    x_train, x_test, y_train, y_test = train_test_split(
        x_data, y_data, test_size=0.1, shuffle=True, random_state=1)
    clf.fit(x_train,y_train)
    y_pred_rf = clf.predict(x_test)   
    acq1 = clf.score(x_test,y_test)
    
#Discretisation parameter. See Fig.6
delta=0.1 

data_analysis(17,train_data,test_data,delta) # Col 18 : Transmission
data_analysis(18,train_data,test_data,delta) # Col 19 : Sx on 6th site
data_analysis(19,train_data,test_data,delta) # Col 20 : Sy on 6th site


# Ploting the Prediction; generates Fig. 4 in the article
ndat=201 #data per set in Test.npy
data=np.load("col17.npy")
for n in [1,2,3]:
    plt.subplot(3,1,n)
    plt.plot(data[(n-1)*ndat:n*ndat,0],data[(n-1)*ndat:n*ndat,2],".",label="Pred") #predicted value
    plt.plot(data[(n-1)*ndat:n*ndat,0],data[(n-1)*ndat:n*ndat,1],"-",label="Calc") #calculated value
plt.legend()
plt.show()

\end{lstlisting}

For quantum machine learning with QSVM we prepare the sample training and test inputs from $TrainQ.npy$. First 2 coloumns of the dataset represent Rashba parameter ($t_R$) and the transmission energy ($E$) respectively. The third column onward represent different output-columns, the sign of non-equilibrium $S_{x,y}$ on each site as the two output classes. In the following we present a sample code for implementing QSVM described in section \ref{subsec: Application of quantum classifier}. 

\begin{lstlisting}[language=Python]
from qiskit import BasicAer
from qiskit.circuit.library import ZZFeatureMap, PauliFeatureMap,ZFeatureMap
from qiskit.aqua import QuantumInstance, aqua_globals
from qiskit.aqua.algorithms import QSVM
from qiskit.aqua.components.multiclass_extensions import AllPairs
from qiskit.aqua.utils.dataset_helper import get_feature_dimension

seed = 10599
aqua_globals.random_seed = 10598
backend = BasicAer.get_backend('statevector_simulator')
quantum_instance = QuantumInstance(backend, shots=1024, seed_simulator=seed, seed_transpiler=seed)
data_map = lambda x: x[0]
feature_map = ZFeatureMap(feature_dimension=get_feature_dimension(training_input), reps=4, data_map_func=data_map)
svm = QSVM(feature_map, training_input, test_input, total_array,multiclass_extension=AllPairs())
result = svm.run(quantum_instance)
for k,v in result.items():
    print(f'{k} : {v}')
\end{lstlisting}


\end{document}